\documentclass[preprint,showpacs,preprintnumbers,amsmath,amssymb]{revtex4-1}

\usepackage{graphicx}
\usepackage{amssymb,amsfonts,amsmath}
\usepackage{color}

\begin{document}

\title{The scaling of human interactions with \mbox{city size}}
\author{Markus Schl\"apfer$^{1,2}$, Lu\'is M. A. Bettencourt$^2$, S\'ebastian Grauwin$^1$, \\
Mathias Raschke$^3$, Rob Claxton$^4$, Zbigniew Smoreda$^5$, Geoffrey B. West$^2$ and Carlo Ratti$^1$}

\affiliation{$^1$Senseable City Laboratory, Massachusetts Institute of Technology, Cambridge, MA, USA\\
$^2$Santa Fe Institute, Santa Fe, NM, USA\\
$^3$Raschke Software Engineering, Wiesbaden, Germany\\
$^4$British Telecommunications plc, Ipswich, UK\\
$^5$Orange Labs, Issy-les-Moulineaux, France}

\begin{abstract}

The size of cities is known to play a fundamental role in social and economic life. Yet, its relation to the structure of the underlying network of human interactions has not been investigated empirically in detail. In this paper, we map society-wide communication networks to the urban areas of two European countries. We show that both the total number of contacts and the total communication activity grow superlinearly with city population size, according to well-defined scaling relations and resulting from a multiplicative increase that affects most citizens. Perhaps surprisingly, however, the probability that an individual's contacts are also connected with each other remains largely unaffected. These empirical results predict a systematic and scale-invariant acceleration of interaction-based spreading phenomena as cities get bigger, which is numerically confirmed by applying epidemiological models to the studied networks. Our findings should provide a microscopic basis  towards understanding the superlinear increase of different socioeconomic quantities with city size, that applies to almost all urban systems and includes, for instance, the creation of new inventions or the prevalence of certain contagious diseases.

\end{abstract}

\maketitle

\section{Introduction}

\vspace{-0.3cm}

The statistical relationship between the size of cities and the structure of the network of human interactions at both the individual and population level has so far not been studied empirically in detail. Early 20th century writings suggested that the social life of individuals in larger cities is more fragmented and impersonal than in smaller ones, potentially leading to negative effects such as social disintegration, crime, and the development of a number of adverse psychological conditions \cite{Simmel:1964, Wirth:1938}. Although some echoes of this early literature persist today, research since the 1970s has dispelled many of these assumptions by mapping social relations across different places \cite{Fischer:1982,Wellman:1999}, yet without providing a comprehensive statistical picture of urban social networks. At the population level, quantitative evidence from many empirical studies points to a systematic acceleration of social and economic life with city size \cite{Milgram:1970, Bornstein:1976}. These gains apply to a wide variety of socioeconomic quantities, including economic output, wages, patents, violent crime and the prevalence of certain contagious diseases ~\cite{Fujita:2001,Sveikauskas:1975,Cullen:1999,CDC:2012}. The average increase in these urban quantities, $Y$, in relation to the city population size, $N$, is well described by superlinear scale-invariant laws of the form $Y \propto  N^\beta$, with a common exponent \mbox{$\beta \approx 1.15>1$ \cite{Bettencourt:2007, Bettencourt:2013}}. 
\vspace{-0.1cm}

Recent theoretical work suggests that the origin of this superlinear scaling pattern stems directly from the network of human interactions \cite{Arbesman:2009, Pan:2013,Bettencourt:2013} - in particular from a similar, scale-invariant increase in social connectivity per capita with city size~\cite{Bettencourt:2013}. This is motivated by the fact that human interactions underlie many diverse social phenomena such as the generation of wealth, innovation, crime or the spread of diseases ~\cite{Anderson:1991, Rogers:1995, Topa:2001, Eubank:2004}. Such conjectures have not yet been tested empirically, mainly because the measurement of human interaction networks across cities of varying sizes has proven to be difficult to carry out. Traditional methods for capturing social networks - for example through \mbox{surveys -} are time-consuming, necessarily limited in scope, and subject to potential sampling biases \cite{Berk:1983}. However, the recent availability of many new large-scale data sets, such as those automatically collected from mobile phone networks \cite{Lazer:2009}, opens up unprecedented possibilities for the systematic study of the urban social dynamics and organisation. 
\vspace{-0.1cm}

In this paper, we explore the relation between city size and the structure of human interaction networks by analysing nationwide communication records in Portugal and the UK. The Portugal data set contains millions of mobile phone call records collected during 15 months, resulting in an interaction network of $1.6 \times 10^6$ nodes and $6.8 \times 10^6$ links (reciprocated social ties). In accordance with previous studies on mobile phone \mbox{networks \cite{Onnela:2007,Raeder:2011, Karsai:2011, Miritello:2013}}, we assume that these nodes represent individuals (subscriptions that indicate business usage are not considered, see Material and Methods). Mobile phone communication data are not necessarily a direct representation of the underlying social network. For instance, two individuals may maintain a strong tie through face-to-face interactions or other means of communication, without relying on regular phone calls \cite{Raeder:2011}. Nevertheless, despite such a potential bias, a recent comparison with a questionnaire-based survey has shown that mobile phone communication data are, in general, a reliable proxy for the strength of individual-based social interactions \cite{Saramaki:2014}. Moreover, even if two subscribers maintain a close relationship and usually communicate via other means, it seems reasonable to assume that both individuals have called each other at least once during the relatively long observation period of 15 months, thus reducing the chance of missing such relationships in our network \cite{Licoppe:2005,Onnela:2007, Krings:2012}. The UK data set covers most national landline calls during 1 month and  the inferred network has $24 \times 10^6$ nodes (landline phones) and $119 \times 10^6$ links, including reciprocated ties to mobile phones (see Material and Methods). We do not consider these nodes as individuals, because we assume that landline phones support the sharing of a single device by several family members or business colleagues \cite{Geser:2006, Onnela:2007}. Nevertheless, conclusions for the total \mbox{(i.e., comprising} the entire population of a city) social connectivity can be drawn. 
\vspace{-0.1cm}

With respect to Portugal's mobile phone data we demonstrate first, that this individual-based interaction network densifies with city size, as the total number of contacts and the total communication activity (call volume and number of calls) grow superlinearly in the number of urban dwellers, in agreement with theoretical predictions and resulting from a continuous shift in the individual-based distributions. Second, we show that the probability that an individual's contacts are also connected with each other (local clustering of links) remains largely constant, which indicates that individuals tend to form tight-knit communities in both small towns and large cities. Third, we show that the empirically observed network densification under constant clustering substantially facilitates interaction-based spreading processes as cities get bigger, supporting the  assumption that the increasing social connectivity underlies the superlinear scaling of certain socioeconomic quantities with city size. Additionally, the UK data suggest that the superlinear scaling of the total social connectivity holds for both different means of communication and different national urban systems.

\section{Results}

\subsection{Superlinear scaling of social connectivity} 

For each city in Portugal, we measured the social connectivity in terms of the total number of mobile phone contacts and the total communication activity (call volume and number of calls). 
Figure 1{\it a} shows the total number of contacts (cumulative degree), $K= \sum_{i \in S} k_i$,  for each Portuguese city  (defined as Statistical City, Larger Urban Zone or Municipality, see Material and Methods) versus its population size, $N$. Here, $k_i$ is the number of individual $i$'s contacts (nodal degree) and $S$  is the set of nodes assigned to a given city. The variation in $K$  is large, even between cities of similar size, so that a mathematical relationship between $K$  and $N$  is difficult to characterise. However, most of this variation is likely due to the uneven distribution of the telecommunication provider's market share, which for each city can be estimated by the coverage  \mbox{$s=\vert S \vert/N$}, with $\vert S \vert$ being the number of nodes in a given city. While there are large fluctuations in the values of $s$, we do not find a statistically significant trend with city size that is consistent across all urban units (see the electronic supplementary material). Indeed, rescaling the cumulative degree by $s$,  $K_r=K/s$, substantially reduces its variation (figure 1{\it b}). Note that this rescaling corresponds to an extrapolation of the observed average nodal degree, $\langle k \rangle = K/\vert S \vert=K_r/N$, to the entire city population. Importantly, the relationship between  $K_r$ and $N$  is now well characterised by a simple power law, $ K_r  \propto N^{\beta}$, with exponent \mbox{$\beta=1.12>1$} (95\% confidence interval (CI) [1.11,1.14]). This superlinear scaling holds over several orders of magnitude and its exponent is in excellent agreement with that of most urban socioeconomic indicators~\cite{Bettencourt:2007}  and with theoretical predictions~\cite{Bettencourt:2013}. The small excess of  $\beta$ above unity implies a substantial increase in the level of social interaction with city size: every doubling of a city's population results, on average, in approximately 12\% more mobile phone contacts per person, as $\langle k \rangle \propto N^{\beta-1}$ with $\beta-1\approx 0.12$. This implies that during the observation period (15 months) an average urban dweller in Lisbon (Statistical City, $N=5\times 10^5$) accumulated about twice as many reciprocated contacts as an average resident of Lixa, a rural town (Statistical City, $N=4 \times10^3$, see \mbox{figure 1{\it c}}). Superlinear scaling with similar values of the exponents also characterises both the population dependence of the rescaled cumulative call volume, $V_r= \sum_{i \in S} v_i/s$, where $v_i$  is the accumulated time user $i$  spent on the phone, and of the rescaled cumulative number of calls, $W_r= \sum_{i \in S} w_i/s$, where  $w_i$ denotes the accumulated number of calls initiated or received by user  $i$, see \mbox{table 1}. Together, the similar values of the scaling exponents for both the number of contacts ($K_r$) and the communication activity ($V_r$ and $W_r$) also suggest, that city size is a less important factor for the weights of links in terms of the call volume and number of calls between each pair of callers. Other city definitions and shorter observation periods \cite{Krings:2012} lead to similar results with overall $\beta=1.05-1.15$ \mbox{(95\% CI [1.00,1.20])}. The non-reciprocal network (see Material and Methods) shows larger scaling exponents $\beta=1.13-1.24$ (95\% CI [1.05,1.25]), suggesting that the number of social solicitations grows even faster with city size than reciprocated contacts. Our predictions for the complete mobile phone coverage are, of course, limited as we only observe a sample of the overall network ($\langle s \rangle \approx 20\%$ for all Statistical Cities, see Material and Methods). Nevertheless, based on the fact that the superlinear scaling also holds when considering only better sampled cities with high values of $s$ (see the electronic supplementary material), and that there is no clear trend in $s$ with city size (so that potential sampling effects presumably apply to urban units of all sizes), we expect that the observed qualitative behaviour also applies to the full network.

For the UK network, despite the relatively short observation period of 31 days, the scaling of reciprocal connectivity shows exponents in the range $\beta=1.08-1.14$ (95\% CI [1.05,1.17]), see table 1. As landline phones may be shared by several people, they do not necessarily reflect an individual-based network and the meaning of the average degree per device becomes limited. Therefore, and considering that the underlying data covers more than 95\% of all residential and business landlines (see Material and Methods), we did not rescale the interaction indicators. Nevertheless, the power law exponents for $K$, $V$ and $W$ (table 1) support the superlinear scaling of the total social connectivity consistent with Portugal's individual-based network, and suggest that this result applies to both different means of communication and different national urban systems.

\subsection{Probability distributions for individual social connectivity}

Previous studies of urban scaling have been limited to aggregated, city-wide quantities~\cite{Bettencourt:2007}, mainly due to limitations in the availability and analysis of extensive individual-based data covering entire urban systems. Here, we leverage the granularity of our data to explore how scaling relations emerge from the underlying distributions of network \mbox{properties}. We focus on Portugal as, in comparison to landlines, mobile phone communication provides a more direct proxy for person-to-person interactions \cite{Eagle:2009, Wesolowski:2013, Saramaki:2014} and is generally known to correlate well with other means of communication \cite{Onnela:2007} and face-to-face meetings \cite{Calabrese:2011}. Moreover, for this part of our analysis we considered only regularly active callers who initiated and received at least one call during each successive period of 3 months, so as to avoid a potential bias towards longer periods of inactivity (see the electronic supplementary material). The resulting statistical distributions of the nodal degree, call volume and number of calls are remarkably regular across diverse urban settings, with a clear shift towards higher values with increasing city size (figure 2). 

To estimate the type of parametric probability distribution that best describes these data, we selected as trial models ({\it i}) the lognormal distribution, ({\it ii}) the generalised Pareto distribution, ({\it iii}) the double Pareto-lognormal distribution and ({\it iv}) the skewed lognormal distribution (see the electronic supplementary material). We first calculated for each interaction indicator, each model $i$ and individual city $c$ the maximum value of the log-likelihood function $\ln L_{i,c}$ \cite{Davidson:2003}. We then deployed it to quantify the Bayesian Information Criterion (BIC) as \mbox{${\rm BIC}_{i,c} =  -2 \ln    L_{i,c} + \eta_i \vert S_c \vert$}, where  $\eta_i$ is the number of parameters used in model $i$  and $\vert S_c \vert$  is the sample size (number of callers in city $c$). The model with the lowest BIC is selected as the best model (see the electronic supplementary material, tables S7-S9). We find that the statistics of the nodal degree is well described by a skewed lognormal distribution (i.e., $k^*=\ln k$  follows a skew-normal distribution), while both the call volume and the number of calls are well approximated by a conventional lognormal distribution (i.e.,  $v^*=\ln v$  and $w^*=\ln w$  follow a Gaussian distribution). The mean values of all logarithmic variables are consistently increasing with city size (figure 2, insets). While there are some trends in the standard deviations (e.g., the standard deviation of $k^*$ is slightly increasing for the Municipalities and the standard deviation of $v^*$ is decreasing for the Statistical Cities), overall we do not observe a clear behaviour consistent across all city definitions. This indicates that superlinear scaling is not simply due to the dominant effect of a few individuals (as in a power-law distribution), but results from an increase in the individual connectivity that characterises most callers in the city. 

More generally, lognormal distributions typically appear as the limit of many random multiplicative processes \cite{Mitzenmacher:2004}, suggesting that an adequate model for the generation of new acquaintances would need to consider a stochastic cascade of new social encounters in space and time that is facilitated in larger cities. As for the analysis of the city-wide quantities (section I.A), the average coverage of $\langle s \rangle \approx 20\%$ may limit our prediction for the complete communication network due to potential sampling effects \cite{Stumpf:2005, Lee:2006}. However, as the basic shape of the distributions is preserved even for those cities with a very high coverage (see the electronic supplementary material, figure S6), we hypothesise that the observed qualitative behaviour also holds for $\langle s \rangle \approx 100\%$.

\subsection{Invariance of the average clustering coefficient}

Finally, we examined the local clustering coefficient, $C_i$, which measures the fraction of connections between one's social contacts to all possible connections between them \cite{Watts:1998}; that is \mbox{$C_i \equiv 2 z_i / [ k_i (k_i-1)]$}, where  $z_i$ is the total number of links between the $k_i$  neighbours of node $i$. A high value of  $C_i$ (close to unity) indicates that most of one's contacts also know each other, while if $C_i=0$ they are mutual strangers. As larger cities provide a larger pool from which contacts can be selected, the probability that two contacts are also mutually connected would decrease rapidly if they were established at random (see the electronic supplementary material). In contrast to this expectation, we find that the clustering coefficient averaged over all nodes in a given city, \mbox{$\langle C \rangle = \sum_{i \in S} C_i/\vert S \vert$}, remains approximately constant with $\langle C \rangle \approx 0.25$ in the individual-based network in Portugal (figure~1{\it c} and figure~3). Moreover, the clustering remains largely unaffected by city size, even when taking into account the link weights (call volume and number of calls, see the electronic supplementary material). The fact that we only observe a sample of the overall mobile phone network in Portugal may have an influence on the absolute value of $\langle C \rangle$ \cite{Lee:2006}, especially if tight social groups may prefer using the same telecommunication provider. Nevertheless, we expect that this potential bias has no effect on the invariance of $\langle C \rangle$, as we do not find a clear trend in the coverage $s$ with city size (see the electronic supplementary material). Thus, assuming that the analysed mobile phone data are a reliable proxy for the strength of social relations \cite{Saramaki:2014}, the constancy of the average clustering coefficient with city size indicates, perhaps surprisingly, that urban social networks retain much of their local structures as cities grow, while reaching further into larger populations. In this context, it is worth noting that  the mobile phone network in Portugal exhibits assortative degree-degree correlations, denoting the tendency of a node to connect to other nodes with similar degree \cite{Raschke:2010} (see the electronic supplementary material). The presence of assortative degree-degree correlations in networks is known to allow high levels of clustering  \cite{Serrano:2005}.

\subsection{Acceleration of spreading processes} 

The empirical quantities analysed so far are topological key factors for the efficiency of network-based spreading processes, such as the diffusion of information and ideas or the transmission of diseases \cite{Boccaletti:2006}. The degree and communication activity (call volume and number of calls) indicate how fast the state of a node may spread to nearby nodes~\cite{Anderson:1991, Satorras:2001, Kitsak:2010}, while the clustering largely determines its probability of propagating beyond the immediate neighbours~\cite{Granovetter:1973, Newman:2009}.  Hence, considering the invariance of the link clustering, the connectivity increase (table 1) suggests that individuals living in larger cities tend to have similar, scale-invariant gains in their spreading potential compared to those living in smaller towns. Given the continuous shift of the underlying distributions \mbox{(figure 2)}, this increasing influence seems to involve most urban dwellers. However, several non-trivial network effects such as community structures \cite{Karsai:2011} or assortative mixing by degree \cite{Kiss:2008} may additionally play a crucial role in the resulting spreading dynamics. 

Thus, to directly test whether the increasing connectivity implies an acceleration of spreading processes, we applied a simple epidemiological model to Portugal's individual-based mobile phone network. The model has been introduced in ref. \cite{Onnela:2007} for the analysis of information propagation through mobile phone communication,  and is similar to the widely used susceptible-infected (SI) model in which the nodes are either in a susceptible or infected state \cite{Anderson:1991}. The spreading is captured by the dynamic state variable $\xi_i(t) \in \{0,1\}$ assigned to each node $i$, with $\xi_i(t) = 1$ if the node is infected (or informed) and $\xi_i(t) = 0$ otherwise. For a given city $c$ we set at time $t=0$ the state of a randomly selected node $i\in S_c$ to $s_i(0)=1$, while all other nodes are in the susceptible (or not-informed) state. At each subsequent time step, an infected node $i$ can pass the information on to each susceptible nearest neighbour $j$ with probability $P_{ij}=x\nu_{ij}$, where $\nu_{ij}$ is the weight of the link between node $i$ and node $j$ in terms of the accumulated call volume, and the parameter $x$ determines the overall spreading speed. Hence, the chance that two individuals will communicate the information increases with the accumulated time they spend on the phone. In accordance with ref. \cite{Onnela:2007}, we choose $x = 1/\nu_{0.9}=1/6242s^{-1}$, with $\nu_{0.9}$ being the value below which 90\% of all link weights in the network fall. This threshold allows to reduce the problem of long simulation running times due to the broad distribution of the link weights, while $P_{ij} \propto \nu_{ij}$ holds for 90\% of all links in the network. The propagation is always realised for the strongest 10\% of the \mbox{links  ($P_{ij}=1$, see \cite{Onnela:2007})}. For each simulation run $\kappa$ we measured  the time $t_{c,\kappa}(n_I)$ until $n_I = \sum_{i \in S_c}\xi_i(t)$ nodes in the given city were infected and estimated the spreading speed as $R_{c,\kappa} =  n_I/ t_{c,\kappa}(n_I)$. The average spreading speed for city $c$ is then given by averaging over all simulation runs, $R_c = \langle R_{c,\kappa} \rangle$. The spreading paths are not restricted to city boundaries but may involve the entire nationwide network. We set the total number of infected nodes to  $n_I=100$ and discarded 4 Statistical Cities and 17 Municipalities for which $\vert S \vert < n_I$. Examples for the infection dynamics and the distribution of the spreading speed resulting from single runs are provided in the electronic supplementary material, figure S10. Figure~4 depicts the resulting values of $R$ for all cities. Indeed, we find a systematic increase of the spreading speed with city size, that can  again be approximated by a power-law scaling relation, $R \propto N^{\delta}$, with $\delta  = 0.11-0.15$ (95\% CI [0.02, 0.26]). Similar increases are also found for simulations performed on the unweighted network (see the electronic supplementary material, figure~S11). These numerical results thus confirm the expected acceleration of spreading processes with city size, and are also in line with a recent simulation study on synthetic networks \cite{Pan:2013}. Moreover, such an increase in the spreading speed has been considered to be a key ingredient for the explanation of the superlinear scaling of certain socioeconomic quantities with city size \cite{Pan:2013,Bettencourt:2013} as, for instance, rapid information diffusion and the efficient exchange of ideas over person-to-person networks can be linked to innovation and productivity \cite{Granovetter:2005,Bettencourt:2013}.

\section{Discussion}

By mapping society-wide communication networks to the urban areas of two European countries, we were able to empirically test the hypothesised scale-invariant increase of human interactions with city size. The observed increase is substantial and takes place within well-defined behavioural constraints in that $i$) the total number of contacts (degree) and the total communication activity (call volume, number of calls) obey superlinear power-law scaling in agreement with theory \cite{Bettencourt:2013} and resulting from a multiplicative increase that affects most citizens, while $ii$) the average local clustering coefficient does not change with city size. Assuming that the analysed data are a reasonable proxy for the strength of the underlying social relations \cite{Saramaki:2014}, and that our results apply to the complete interaction networks, the constant clustering is particularly noteworthy as it suggests that even in large cities we live in groups that are as tightly knit as those in small towns or \mbox{`villages' \cite{Jacobs:1961}}. However, in a real village we may need to accept a community imposed on us by sheer proximity, whereas in a city we can follow the homophilic tendency \cite{McPherson:2001} of choosing our own village - people with shared interests, profession, ethnicity, sexual orientation, etc. Together, these characteristics of the analysed communication networks indicate that  larger cities may facilitate the diffusion of information and ideas or other interaction-based spreading processes. This further supports the prevailing hypothesis that the structure of social networks underlies the generic properties of cities, manifested in the superlinear scaling of almost all socioeconomic quantities with population size. 

The wider generality of our results remains, of course, to be tested on other individual-based communication data, ideally with complete coverage of the population  \mbox{($\langle s \rangle \approx 100\%$)}. Nevertheless, the revealed patterns offer a baseline to additionally explore the differences of particular cities with similar size, to compare the observed network properties with face-to-face interactions \cite{Calabrese:2011} and to extend our study to other cultures and economies. Furthermore, it would be instructive to analyse in greater detail how cities affect more specific circles of social contacts such as family, friends or business colleagues \cite{Saramaki:2014, Miritello:2013}. Finally, it remains a challenge for future studies to establish the causal relationship between social connectivity at the individual and organisational levels and the socioeconomic characteristics of cities, such as economic output, the rate of new innovations, crime or the prevalence of contagious diseases. To that end, in combination with other socioeconomic or health-related data, our findings might serve as a microscopic and statistical basis for network-based interaction models in sociology~\cite{Wasserman:1994, Lazer:2009}, economics~\cite{Fujita:2001, Eagle:2010} and epidemiology \cite{Eubank:2004}.
\newpage

\section{Material and methods}
\footnotesize

\subsection{Data sets} 

The Portugal data set consists of 440 million Call Detail Records (CDR) from 2006 and 2007, covering voice calls of $\approx$ 2 million mobile phone users and thus \mbox{$\approx$ 20\%} of the country's population (in 2006 the total mobile phone penetration rate was $\approx$ 100\%, survey available at http://www.anacom.pt). The data has been collected by a single telecom service provider for billing and operational purposes. The overall observation period is 15 months during which the data from 46 consecutive days is lacking, resulting in an effective analysis period of $\Delta T = 409$ days. To safeguard privacy, individual phone numbers were anonymised by the operator and replaced with a unique security ID. Each CDR consists of the IDs of the two connected individuals, the call duration, the date and time of the call initiation, as well as the unique IDs of the two cell towers routing the call at its initiation. In total, there are 6511 cell towers for which the geographic location was provided, each serving on average an area of 14 km$^2$, which reduces to  0.13 km$^2$ in urban areas. The UK data set contains 7.6 billion calls from a one-month period in 2005, involving 44 million landline and 56 million mobile phone numbers ($>95\%$ of all residential and business landlines countrywide). For customer anonymity, each number was replaced with a random, surrogate ID by the operator before providing the data. We had only partial access to the connections made between any two mobile phones. The operator partitioned the country into 5500 exchange areas (covering 49 km$^2$ on average), each of which comprises a set of landline numbers. The data set contains the geographic location of 4000 exchange areas.

\subsection{City definitions}

Because there is no unambiguous definition of a city we explored different units of analysis. For Portugal, we used the following city definitions: (\textit{i}) Statistical Cities (STC), (\textit{ii}) Municipalities (MUN) and (\textit{iii}) Larger Urban Zones (LUZ). STC and MUN are defined by the Portuguese National Statistics Office (http://www.ine.pt), which provided us with the 2001 population data, and with the city perimeters (shapefiles containing spatial polygons). The LUZ are defined by the European Union Statistical Agency (ÔEurostatÕ) and correspond to extended urban regions (the population statistics and shapefiles are publicly available at http://www.urbanaudit.org). For the LUZ, we compiled the population data for 2001 to assure comparability with the STC and MUN. In total, there are 156 STC, 308 MUN and 9 LUZ. The MUN are an administrative subdivision and partition the entire national territory. Although their interpretation as urban units is flawed in some cases, the MUN were included in the study as they cover the total resident population of Portugal. There are 6 MUN which correspond to a STC. For the UK, we focussed on Urban Audit Cities (UAC) as defined by Eurostat, being equivalent to Local Administrative Units, Level 1 \mbox{(LAU-1)}. Thus, using population statistics for 2001 allows for a direct comparison with the MUN in Portugal (corresponding to LAU-1). In total, the UK contains 30 UAC.

\subsection{Spatial interaction networks}

For Portugal, we inferred two distinct types of interaction networks from the CDRs: in the reciprocal (REC) network each node represents a mobile phone user and two nodes are connected by an undirected link if each of the two corresponding users initiated at least one call to the other. In accordance with previous studies on mobile phone data \cite{Onnela:2007, Miritello:2013}, this restriction to reciprocated links avoids subscriptions that indicate business usage (large number of calls which are never returned) and should largely eliminate call centers or accidental calls to wrong subscribers. In the non-reciprocal (nREC) network two nodes are connected if there has been at least one call between them. The nREC network thus contains one-way calls which were never reciprocated, presumably representing more superficial interactions between individuals which might not know each other personally. Nevertheless, we eliminated all nodes which never received or never initiated any call, so as to avoid a potential bias induced by call centres and other business hubs. We performed our study on the largest connected cluster (LCC, corresponding to the giant weakly connected component for the nREC network) extracted from both network types (see the electronic supplementary material, table S1). In order to assign a given user to one of the different cities, we first determined the cell tower which routed most of his/her calls, presumably representing his or her home place. Subsequently, the corresponding geographic coordinate pairs were mapped to the polygons (shapefiles) of the different cities. Following this assignment procedure, we were left with 140 STC (we discarded 5 STC for which no shapefile was available and 11 STC without any assigned cell tower), 9 LUZ and 293 MUN (we discarded 15 MUN without any assigned cell tower), see the electronic supplementary material, figure S1 and table S2, for the population statistics. The number of assigned nodes is strongly correlated with city population size ($r$=0.95,0.97,0.92 for STC, LUZ and MUN, respectively, with p-value$<$0.0001 for the different urban units), confirming the validity of the applied assignment procedure. To further test the robustness of our results, we additionally determined the home cell tower by considering only those calls that were initiated between 10pm and 7am, yielding qualitatively similar findings to those reported in the main text. For the UK, due to limited access to calls among mobile phones and to insufficient information about their spatial location, we included only those mobile phone numbers that had at least one connection to a landline phone. Subsequently, in order to reduce a potential bias induced by business hubs, we followed the data filtering procedure used in \cite{Eagle:2010}. Hence, we considered only the REC network and we excluded all nodes with a degree larger than 50, as well as all links with a call volume exceeding the maximum value observed for those links involving mobile phone users. Summary statistics are given in the electronic supplementary material, table S3. We then assigned an exchange area together with its set of landline numbers to an UAC, if the centre point of the former is located within the polygon of the latter. This results in 24 UAC containing at least one exchange area (see the electronic supplementary material, figure S2 and table S4).

\begin{acknowledgments}
We thank Jos\'e Lobo, Stanislav Sobolevsky, Michael Szell, Riccardo Campari, Benedikt Gross and Janet Owers for comments and discussions. M.S., S.G. and C.R. gratefully acknowledge British Telecommunications plc, Orange Labs, the National Science Foundation, the AT\&T Foundation, the MIT Smart Program, Ericsson, BBVA, GE, Audi Volkswagen, Ferrovial and the members of the MIT Senseable City Lab Consortium. L.M.A.B. and G.B.W. acknowledge partial support by the Rockefeller Foundation, the James S. McDonnell Foundation (grant no. 220020195), the National Science Foundation (grant no. 103522), the John Templeton Foundation (grant no. 15705) and the  Army Research Office Minerva Program (grant no. W911NF1210097). Mobile phone and landline data were provided by anonymous service providers in Portugal and the UK and are not available for distribution.
\end{acknowledgments}

\begin{figure}
{\includegraphics[width=9cm]{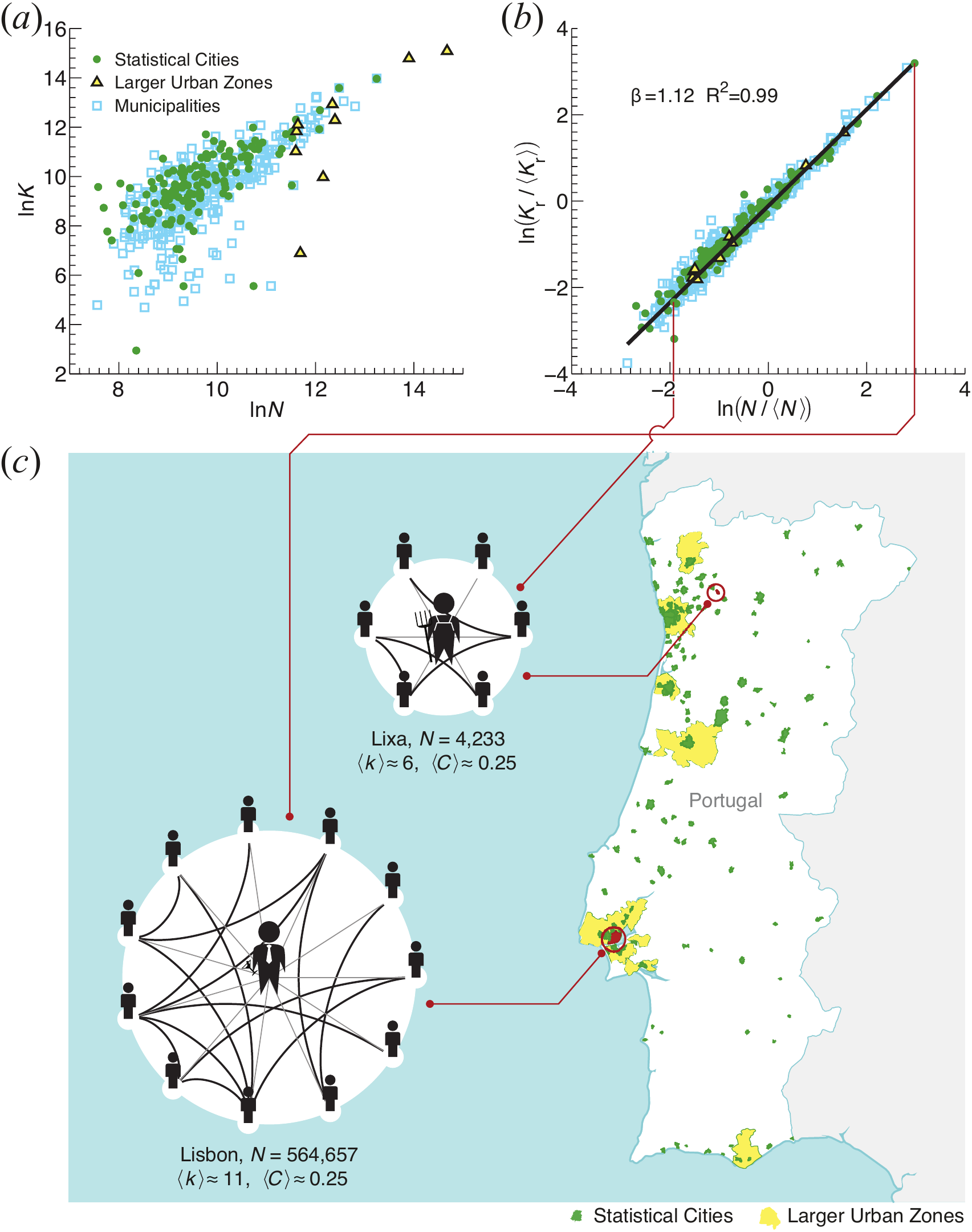}}
\caption{Human interactions scale superlinearly with city size. ({\it a}) Cumulative degree,  $K$, versus city population size, $N$,  for three different city definitions in Portugal. ({\it b}) Collapse of  the cumulative degree onto a single curve after rescaling by the coverage, $K_r=K/s$. For each city definition, the single values of  $K_r$  and $N$  are normalised by their corresponding average values, $\langle K_r \rangle $  and $\langle N \rangle$, for direct comparison across different urban units of analysis. ({\it c}) An average urban dweller of Lisbon has approximately twice as many reciprocated mobile phone contacts, $\langle k \rangle$, than an average individual in the rural town of Lixa. The fraction of mutually interconnected contacts (black lines) remains unaffected, as indicated by the invariance of the average clustering coefficient,  $\langle C \rangle$. The map further depicts the location of Statistical Cities and Larger Urban Zones in Portugal, with the exception of those located on the archipelagos of the Azores and Madeira.}
\label{Fig-1}
\end{figure}

\begin{figure*}
\centerline{\includegraphics[width=14.5cm]{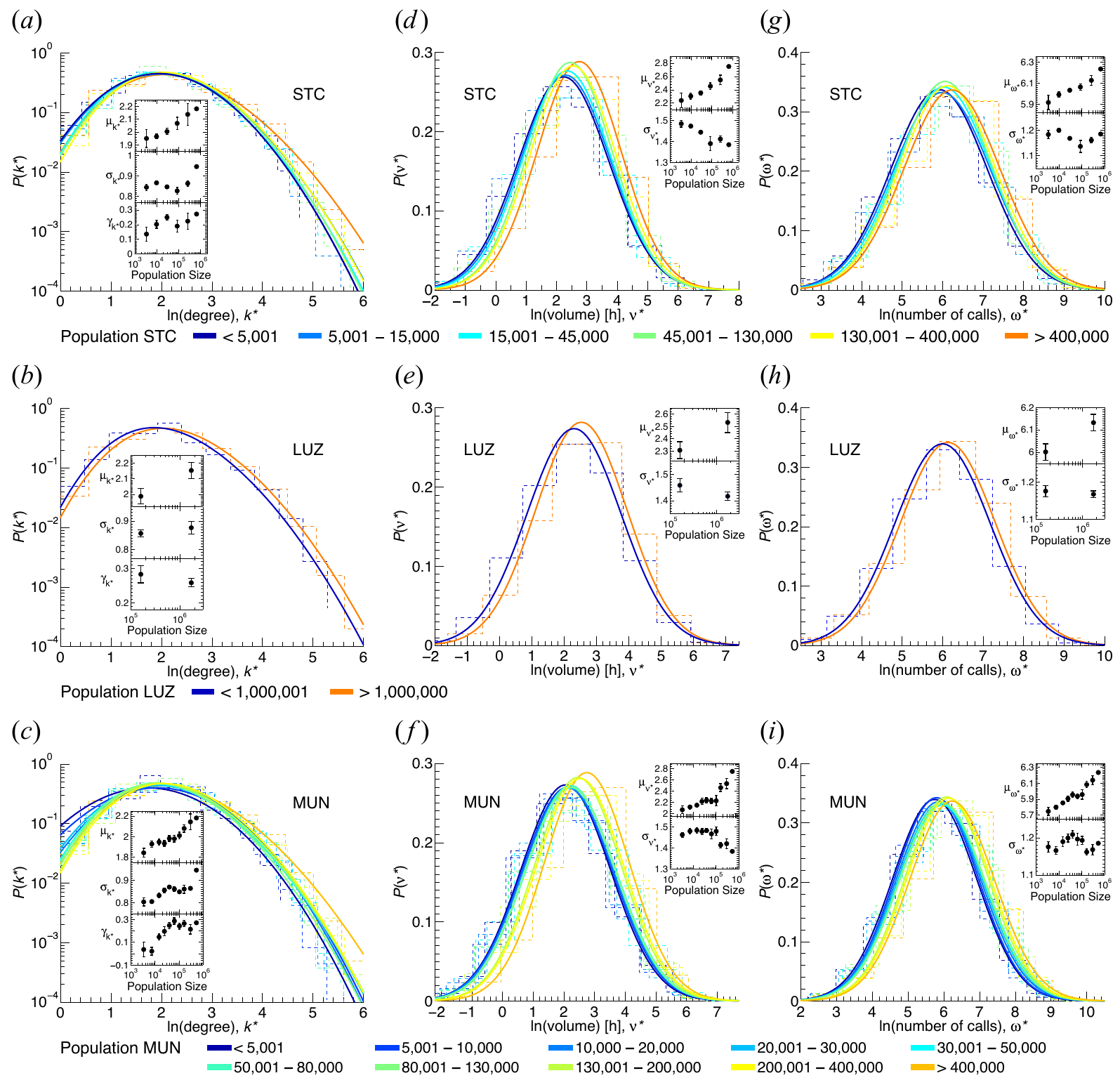}}
\caption{The impact of city size on human interactions at the individual level. ({\it a-c}) Degree distributions, $P(k^*)$, for Statistical Cities (STC), Larger Urban Zones (LUZ) and Municipalities (MUN); the individual urban units are log-binned according to their population size. The dashed lines indicate the underlying histograms and the continuous lines are best fits of the skew-normal distribution with mean  $\mu_{k^*}$, standard deviation $\sigma_{k^*}$  and skewness $\gamma_{k^*}$ (insets). ({\it d-f}) Distributions of the call volume, $P(v^*)$, and ({\it g-i}) number of calls, $P(w^*)$; the continuous lines are best fits of the normal distribution with mean values $\mu_{v^*}$  and $\mu_{w^*}$, and standard deviations $\sigma_{v^*}$ and $\sigma_{w^*}$, respectively (insets). Error bars denote the standard error of the mean (SEM). The distribution parameters are estimated by the maximum likelihood method, see the electronic supplementary material.}
\label{Fig-2}
\end{figure*}

\begin{figure}
{\includegraphics[width=7cm]{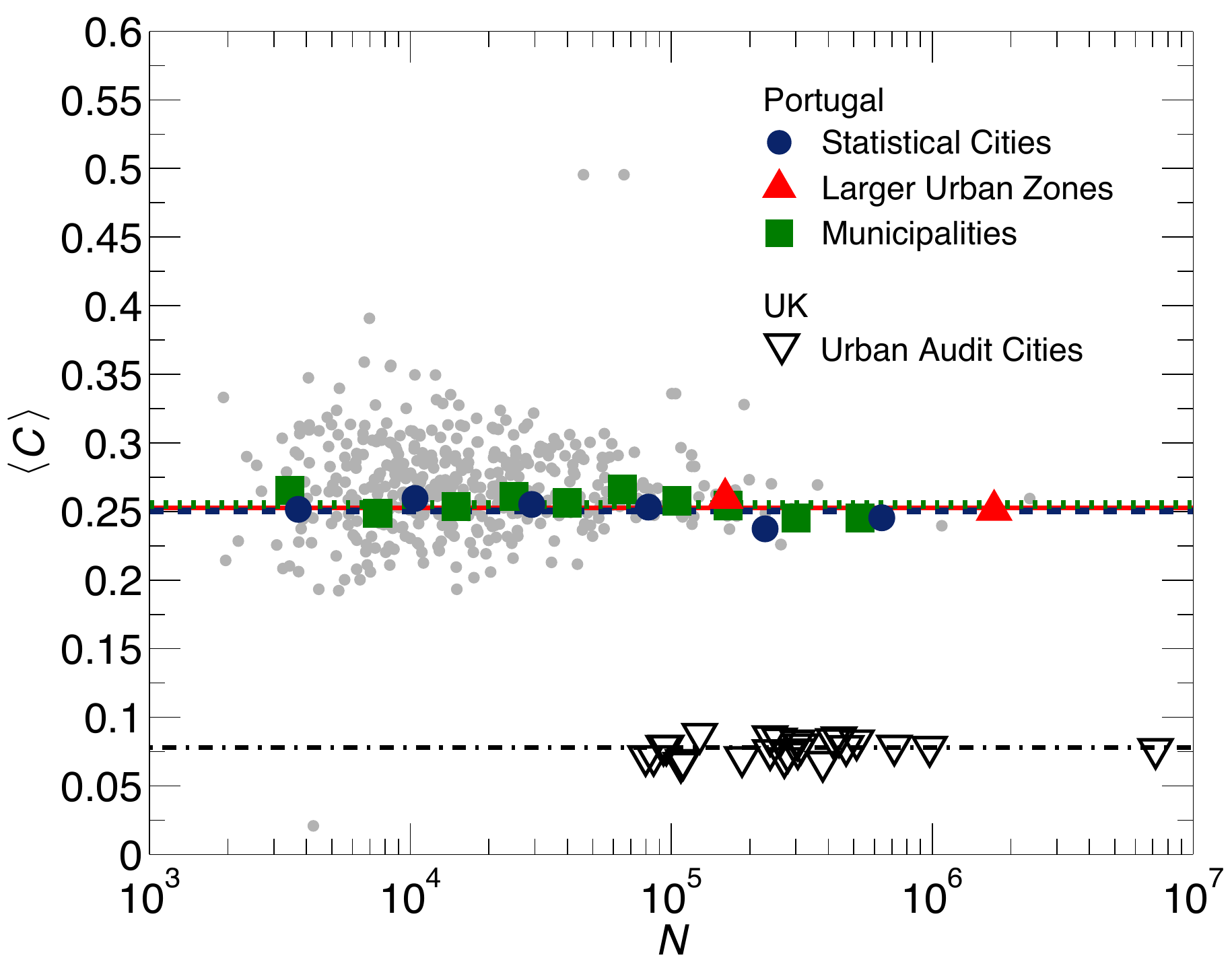}}
\caption{The average clustering coefficient remains unaffected by city size. The lines indicate the average values with $0.251 \pm 0.021$ for STC (weighted average and standard deviation, dashed line), $0.252 \pm 0.013$ for LUZ (continuous line) and $0.255 \pm 0.021$ for MUN (dotted line) in Portugal, and $0.078 \pm 0.004$ for the UK (dash-dotted line). For Portugal, the individual urban units are log-binned according to their population size as in figure~2, to compensate for the varying coverage of the telecommunication provider. The error bars (SEM) are smaller than the symbols. Grey points are the underlying scatter plot for all urban units. A regression analysis on the data is provided in the electronic supplementary material, figure S7. The value of $\langle C \rangle$  in the UK is lower than in Portugal, as expected for a landline network that captures the aggregated activity of different household members or business colleagues. If we assume that an average landline in the UK is used by 3 people who communicate with a separate set of unconnected friends, we would indeed expect that the clustering coefficient would be approximately 1/3 of that of each individual.}
\label{Fig-3}
\end{figure}

\begin{figure}
\centerline{\includegraphics[width=7.2cm]{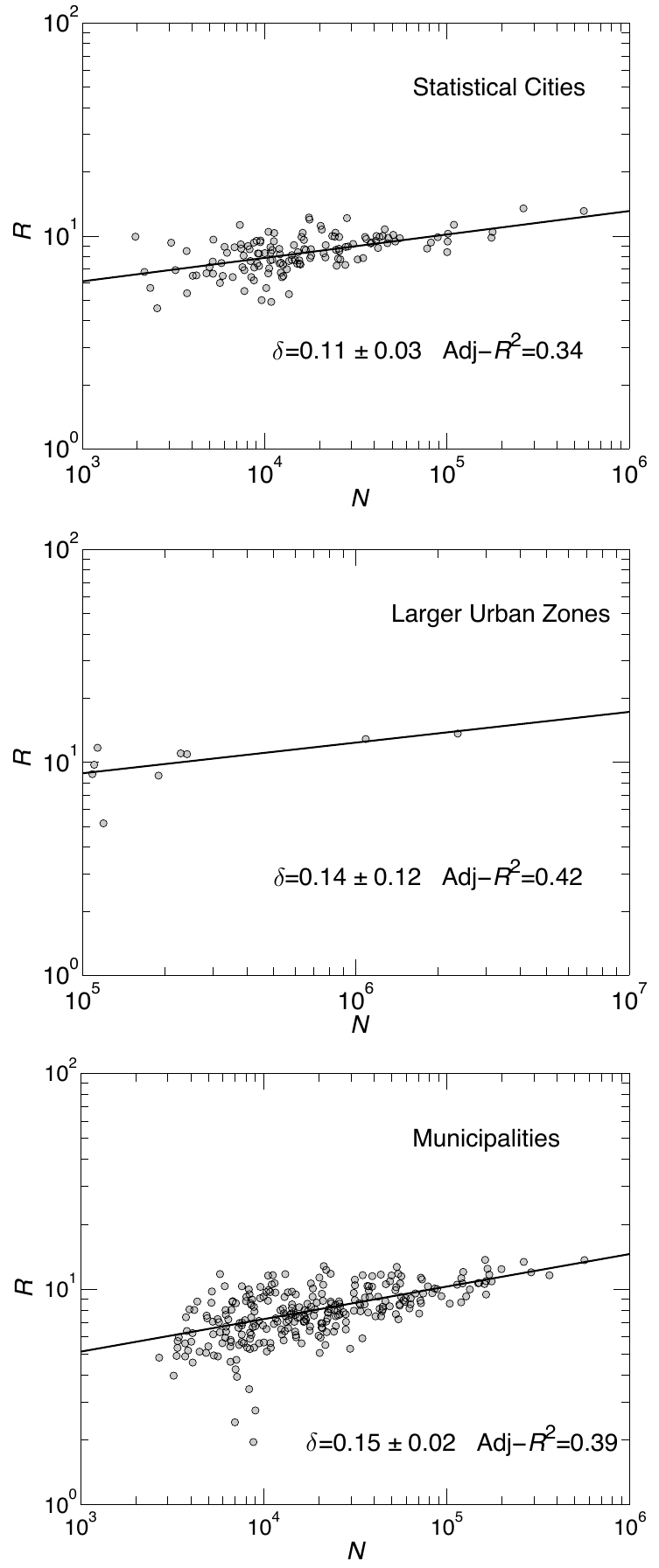}}
\label{Fig-4}
\caption{Larger cities facilitate interaction-based spreading processes. The panels show the average spreading speed versus city size, broken down into the different city definitions. For each urban unit, the values of $R$ result from averaging over 100 simulation trials performed on the reciprocal network in Portugal ($\Delta T = 409$ days), weighted by the accumulated call volume between each pair of nodes. The solid lines are the best fit of a power-law scaling relation $R  \propto N^{\delta}$, for which the values of the exponent, the corresponding 95\% confidence intervals and the coefficients of determination are indicated.}
\end{figure}

\setlength\tabcolsep{3pt}
\begin{table}[h]
\footnotesize
\caption{\footnotesize Scaling exponents $\beta$. The observation period of $\Delta T=409$ days is the full extent of the Portugal data set, while $\Delta T=92$ days corresponds to the first three consecutive months. For the call volume statistics, we discarded 1 Larger Urban Zone (Ponta Delgada) due to a high estimation error of  $V_r$ (SEM $>$ 20\%). For the UK data, the interaction indicators,  $Y$, are not rescaled by the coverage due to consistently high market share. The indicator $K_{lm}$  is based on the cumulative number of links between landlines and mobile phones only (landline-landline connections are excluded). Exponents were estimated by nonlinear least squares regression (trust-region algorithm), with Adj-$R^2>0.98$ for all fits.}
\begin{tabular}{@{\vrule height 5.5pt depth4pt  width0pt}lrcccccc}
&\multicolumn5c{}\\
\noalign{\vskip-11pt}
\vrule depth 6pt width 0pt Portugal & City Definition &Number & Network Type &$ \Delta T$ & $Y$ & $\beta$ & 95\% CI\\
\hline
&Statistical City \parbox[0pt][1.4em]{0cm}&140	& reciprocal &	409 days & Degree ($K_r$)& 1.12&[1.11, 1.14] \\  [-0.5em]

& & & & & Call volume ($V_r$) & 1.11 & [1.09, 1.12] \\[-0.5em]
& & & & & Number of calls ($W_r$) & 1.10 & [1.09, 1.11] \\[0.2em]
& & & & 92 days & Degree ($K_r$) &1.10 & [1.09, 1.11] \\ [-0.5em]
& & & & & Call volume ($V_r$) & 1.10 & [1.08, 1.11] \\ [-0.5em]
& & & & & Number of calls ($W_r$) & 1.08 &[1.07, 1.10] \\[0.2em]
& & &  non-reciprocal &	409 days & Degree ($K_r$) & 1.24 & [1.22, 1.25] \\ [-0.5em]
& & & & &	Call volume ($V_r$) &  1.14 & [1.12, 1.15] \\ [-0.5em]
& & & & &  Number of calls ($W_r$) & 1.13 & [1.12, 1.14] \\[0.2em]
& Larger Urban Zone & 9(8) &	reciprocal	 & 409 days &	Degree ($K_r$) & 1.05 & [1.00, 1.11] \\ [-0.5em]
& & & & &	Call volume ($V_r$) & 1.11 & [1.02, 1.20] \\ [-0.5em]
& & & & &	Number of calls ($W_r$) & 1.10 & [1.05, 1.15] \\[0.2em]
& & & non-reciprocal & 409 days &	Degree ($K_r$) &  1.13 &	[1.08, 1.18] \\ [-0.5em]
& & & & &	 Call volume ($V_r$) & 1.14	& [1.05, 1.23] \\ [-0.5em]
& & & & &	Number of calls ($W_r$) & 1.13 & [1.08, 1.18] \\[0.2em]
& Municipality & 293	 & reciprocal & 409 days & Degree ($K_r$) & 1.13 &	[1.11, 1.14] \\ [-0.5em]
& & & & &	 Call volume ($V_r$) & 1.15	& [1.13, 1.17] \\ [-0.5em]
& & & & &	 Number of calls ($W_r$) & 1.13	& [1.11, 1.14] \\[0.2em]
\vrule depth 6pt width 0pt UK & City Definition &Number & Network Type &$ \Delta T$ & $Y$ & $\beta$ & 95\% CI  \parbox[0pt][2em][c]{0cm}{}\\
\hline
 & Urban Audit City	\parbox[0pt][1.4em]{0cm} & 24	& reciprocal &	31 days	& Degree ($K$) &	1.08	 & [1.05, 1.12] \\ [-0.5em]
 & & & & & Degree, land-mobile ($K_{lm}$) & 1.14 &	[1.11, 1.17] \\ [-0.5em]
 & & & & & Call volume ($V$) &	1.10	& [1.07, 1.14] \\ [-0.5em]
 & & & & & Number of calls ($W$)	& 1.08 &	[1.05, 1.11] \\
\hline
\end{tabular}
\end{table}

\end{document}